\documentclass{article}
\usepackage{amssymb}

\begin{document}
\author{Anatoly Konechny${\,}^{1}$ and  Albert Schwarz${\,}^{2}$\\
 \\
${}^{1}\,$Department of Physics, University of California Berkeley \\
and \\
Theoretical Physics Group, Mail Stop 50A-5101\\
LBNL, Berkeley, CA 94720 USA \\ 
konechny@thsrv.lbl.gov\\
\\
${}^{2}\,$Department of Mathematics, University of California Davis\\
Davis, CA 95616 USA\\
  schwarz@math.ucdavis.edu}

\title{\bf Moduli spaces of maximally supersymmetric  solutions on noncommutative tori and 
noncommutative orbifolds }
\maketitle
\large
\begin{abstract}
\large
A maximally supersymmetric configuration of super Yang-Mills living on a noncommutative
torus corresponds to a constant curvature connection. On a noncommutative toroidal 
orbifold there is an additional constraint that the connection be equivariant. 
We study moduli spaces of (equivariant) constant curvature connections on 
noncommutative even-dimensional tori and on toroidal orbifolds. 
As an illustration we work out the cases of ${\bf Z}_{2}$ and ${\bf Z}_{4}$ orbifolds 
in detail. The results we obtain 
agree with a commutative picture describing   systems of branes wrapped on cycles of the torus and 
branes stuck at exceptional orbifold points.
\end{abstract}
\large
\section{Introduction}
The idea that String theory leads to some sort of fuzzy or noncommutative 
microscopic structure of space-time has been around for quite a while. 
 A new boost to this idea emerged  after  the paper \cite{CDS}. 
It was shown in that paper that noncommutative tori arise as particular compactifications 
of M(atrix) theory (\cite{BFSS}) that is conjectured to be a nonperturbative definition of 
String theory (see \cite{DougHull} - \cite{Schomerus} and references therein for  the subsequent development). 
Later in \cite{SeibWitt} a number of important results concerning relations between String theory 
and noncommutative geometry were obtained. In particular the conditions
under which noncommutative geometry arises within perturbative open string theory were clarified.

On the mathematics side noncommutative tori are the best studied examples of noncommutative spaces 
(see \cite{Rieffel1} for a good overview). An important notion in noncommutative geometry is that of 
Morita equivalence that gives some equivalence relation between algebras of functions on noncommutative spaces. 
A striking result first observed in \cite{CDS} and later on proved rigorously in \cite{ASMorita} is 
that noncommutative world volume field theories living on noncommutative tori are invariant under 
duality transformations generated by Morita equivalences of noncommutative tori. This duality 
is directly related to T-duality of perturbative string theory compactifications  (\cite{ASRieffel}, \cite{SeibWitt}).

It seems to be    interesting and important to study compactifications on other noncommutative spaces and see 
how much of noncommutative geometry  techniques that proved to be useful for tori can be extended (see \cite{HoWu}, \cite{HoWu2}, 
\cite{ncRS}, \cite{z2} for some work done in that direction). 
In paper \cite{z2} 
we  studied  M(atrix) theory compactifications on noncommutative toroidal orbifolds.  
That paper primarily concentrated on 
 two-dimensional  ${\bf Z}_{2}$ orbifolds. 
In the present paper we continue  investigation of  M(atrix) theory compactification on noncommutative spaces.
The main topic of this paper is the structure of projective modules over noncommutative orbifolds that admit a  
constant curvature 
Yang-Mills field  and moduli spaces of all such fields. Such modules and constant curvature connections 
on them describe configurations preserving half of the unbroken supersymmetries. 
 All our results concern classical aspects only. However  counting of quantum states 
with specified brane charges can be made in terms of a supersymmetric sigma model on the appropriate classical 
moduli space provided we have a sufficient number of supersymmetries.

In the commutative case configurations with vanishing $SU(N)$ part of the curvature and their interpretations 
in terms of D-branes were considered in a number of papers (see \cite{GurRamg}  and references 
therein). D0 branes on toroidal ${\bf Z}_{2}$ orbifolds were studied in \cite{RamgWald}, \cite{GLY}. 
Whenever our results can be compared with the commutative results we observe a complete agreement.

The paper is organized as follows. Section 2 contains some generalities on noncommutative tori 
and matrix theory as well as an explanation of our general strategy regarding a moduli space problem. 
In essence the novelty of our approach is in the following. In the usual approach one  first fixes a module, i.e. a 
representation of algebra of functions on a noncommutative space,  
and then considers all constant curvature connections modulo gauge transformations. Instead we fix the connection $\nabla_{i}$ and 
then look at all equivalence classes of the torus representations compatible with that connection. 
If $[\nabla_{i}, \nabla_{j}] = f_{ij}\cdot {\bf 1}$ where $f_{ij}$ is a non-degenerate matrix then the  algebra generated by 
$\nabla_{i}$ is isomorphic to Heisenberg algebra. We can use then the well known results about complete reducibility 
of Heisenberg algebra representations. The case of generic constant curvature connection can be reduced to the nondegenerate 
case by use of Morita equivalence provided the torus dimension is even. The  Morita equivalence technique by itself is not directly 
relevant to the present paper; its discussion is relegated to the appendix. 
We restrict ourselves to the even dimensional 
case in this paper. However after appropriate modification our method also works for odd dimensions.

In section 3 we apply our general strategy to noncommutative tori and prove that the moduli space 
is isomorphic to $(\tilde T^{d})^{r}/S_{r}$ where $\tilde T^{d}$ is a commutative $d$-dimensional torus and 
$r$ is a greatest common divisor of topological integers 
(D-brane charges) characterizing the module. For $d=2$ this result was proved by A.~Connes and M.~Rieffel in \cite{ConnesRieffel}.

In section 4 we introduce noncommutative toroidal orbifolds and  outline how the approach used for 
tori can be extended to the study of moduli space of equivariant constant curvature connections.

In sections 5 and 6 we study in detail ${\bf Z}_{2}$ and ${\bf Z}_{4}$ orbifolds respectively. 
For this cases we work out  a general construction of a module that admits a constant curvature equivariant connection.
In each case a module is built out of standard blocks that can be interpreted to describe D0 particles and 
various membranes stuck at exceptional points of the orbifold. The corresponding moduli spaces are proved to be 
isomorphic to $(\tilde T^{d}/{\bf Z}_{2})^{m})/S_{m}$ and  $(\tilde T^{d}/{\bf Z}_{4})^{n}/S_{n}$ respectively. 
Here $m$ and $n$ are some integers depending on  topological numbers of the module. 
In the two-dimensional case ${\bf Z}_{2}$ and ${\bf Z}_{4}$ orbifolds of  noncommutative tori were 
studied by S.~Walters in a number of papers (\cite{Walters}). It would be interesting to calculate the topological 
numbers introduced in those papers for the modules we consider.

In section 7 we add scalar fields to the discussion and describe how Coulomb branches vary over 
the moduli space of constant curvature connections. 

Finally the appendix contains general discussion of Morita equivalence for toroidal orbifolds and 
details of its application to ${\bf Z}_{2}$ and ${\bf Z}_{4}$ cases. 

\section{ 1/2 BPS configurations on noncommutative tori}
We start this section by discussing  some general aspects 
of constant curvature connections on noncommutative tori. 

We define an algebra $A_{\theta}$ of smooth functions on a $d$-dimensional 
noncommutative torus in the following way. Let $D\subset {\bf R}^{d}$ 
be a lattice $D\cong {\bf Z}^{d}$ and $\theta_{ij}$ be an antisymmetric 
$d\times d$ matrix. The algebra $A_{\theta}$ is an associative algebra 
whose elements are formal series 
$$
\sum_{{\bf n}\in D} C({\bf n})U_{\bf n}
$$ 
where $C({\bf n})$ are complex numbers and $U_{\bf n}$ satisfy the relations 
\begin{equation} \label{nct}
U_{\bf n}U_{\bf m}=U_{\bf n + m }e^{\pi i n_{j}\theta^{jk}m_{k}} \, .
\end{equation}
Here the coefficients $ C({\bf n})$ are assumed to decrease faster than 
any power of $|{\bf n}|$ as ${\bf n}\to \infty$. We will denote 
generators corresponding to standard basis vectors 
$(n^{i})_{j}  = \delta^{i}_{j}$ by $U_{i}$. Any $U_{\bf n}$ can be 
expressed as a product of $U_{i}$'s times a numerical phase factor.   

The notion of a vector bundle (or rather that of a space of sections of 
a vector bundle) in noncommutative geometry is replaced by a projective module 
over $T_{\theta}$. Unless specified we will assumed that we work with a 
right module (i.e. we have a right action of $T_{\theta}$). 
Let $L$ be a $d$-dimensional commutative Lie algebra acting 
on $T_{\theta}$ by means of derivations
\begin{equation}
\delta_{j} U_{\bf n} = 2\pi i n_{j} U_{\bf n} \, .
\end{equation} 
A connection  on a projective module $E$ over $T_{\theta}$ 
is defined in terms of operators $\nabla_{i}: E\to E$ satisfying 
a Leibniz rule 
$$
\nabla_{i}(ea) = (\nabla_{i}e)a + e(\delta_{i}a)
$$ 
for any $e\in E$  and any $a\in T_{\theta}$.

It was first shown in (\cite{CDS}) how (super)Yang-Mills theory 
on a noncommutative torus arises as a particular compactification of 
M(atrix) theory (\cite{BFSS}, \cite{IKKT}). 
The Minkowski action functional of M(atrix) theory compactified on a noncommutative
torus $T_{\theta}$ can be written as
\begin{eqnarray} \label{action}
S =&& \frac{-V}{4g_{YM}^{2}} {\rm Tr} \bigl( (F_{ij} + \phi_{ij}{\bf 1})^{2}
 + [\nabla_{i}, X_{I}]^{2} + [X_{I}, X_{J}]^{2} \bigr) \nonumber \\
&& + \frac{iV}{2g_{YM}^{2}} {\rm Tr} \bigl( \bar \psi \Gamma^{i} [\nabla_{i} ,
\psi ] + \bar \psi \Gamma^{I} [X_{I} ,\psi ] \bigr)
\end{eqnarray}
where  $\nabla_{i}$, $i=1,\dots, d$ is a connection on a $T_{\theta}$-module $E$, 
$\psi$ is a ten-dimensional Majorana-Weyl spinor taking values in the algebra $End_{T_{\theta}}E$ 
of endomorphisms of $E$, $\phi_{ij}$ is an antisymmetric tensor,  $X_{I}\in End_{T_{\theta}}E$, $I=d+1, \dots 10$  
are scalar fields, 
$\Gamma_{\mu}$ are ten-dimensional gamma-matrices. 

The  action (\ref{action}) is invariant
under the following supersymmetry
transformations
\begin{eqnarray} \label{susy}
&&\delta \nabla_{i} = \frac{i}{2}\bar \epsilon \Gamma_{i}\psi
\nonumber \\
&&\delta X_{I} = \frac{i}{2}\bar \epsilon \Gamma_{I}\psi
\nonumber \\
&& \delta \psi = \tilde \epsilon -\frac{1}{4} (F_{ij}\Gamma^{ij} + 2[\nabla_{i}, X_{I}]\Gamma^{iI} + [X_{I}, X_{J}]\Gamma^{IJ})
\epsilon 
\end{eqnarray}
where $\epsilon$, $\tilde \epsilon$ are 10D Majorana-Weyl spinors parameterizing the transformation.

Classical configurations preserving 1/2 of these supersymmetries satisfy the equations
\begin{eqnarray}\label{1/2BPS}
&& [\nabla_{j}, \nabla_{k}]=2\pi i f_{jk}\cdot {\bf 1} \, , \nonumber \\
&& [\nabla_{j}, X_{I}] = 0 \, , \qquad [X_{I}, X_{J}]=0 
\end{eqnarray}
where $f_{ij}$ are constants and $\bf 1$ is a unit endomorphism. 
We call solutions to the first equation modulo gauge transformations  a Higgs branch of the 1/2-BPS moduli space. 
The whole moduli space of solutions to (\ref{1/2BPS}) can be viewed as a fibration over the Higgs branch. 
We will first describe the Higgs branch which is the moduli space of constant curvature connections and then 
take into account the scalar fields.

Let us  outline here the strategy we  will take addressing the moduli space problem. 
The complete set of equations that gives  a module over  a noncommutative torus  and a constant curvature 
connection on it reads 
\begin{eqnarray} \label{eqs}
&&U_{j}U_{k} = e^{2\pi i \theta_{jk}} U_{k}U_{j} \, , \nonumber \\
&&[\nabla_{j}, U_{k} ] = 2\pi i \delta_{jk} U_{k} \, , \nonumber \\
&&  [\nabla_{j}, \nabla_{k}]=2\pi i f_{jk} 
\end{eqnarray}   
If $U_{i}$, $\nabla_{j}$ are operators in Hilbert space $E$ satisfying these equations and 
 $Z: E\to E$ is a unitary linear (over $\bf C$) operator, then $ZU_{i}Z^{-1}$, $Z\nabla_{i}Z^{-1}$ also solve (\ref{eqs}). 
The usual approach to equations  (\ref{eqs}) is to fix generators $U_{i}$, i.e. fix a module, and then look for $\nabla_{i}$ 
satisfying the last two equations in (\ref{eqs}). 
The transformations $Z: E\to E$ preserving $U_{i}$'s are unitary endomorphisms.
We will take a different approach in this paper and fix the representation of $\nabla_{i}$. The gauge transformations in this 
picture are given by unitary operators $Z$ that commute with all $\nabla_{i}$. The moduli space is then a space of 
solutions $U_{i}$ to the first two equations in (\ref{eqs}) modulo gauge transformations. 
Note that in this approach  inequivalent  modules, that in our case are  modules that have distinct Chern (and/or other topological 
numbers in the case of orbifolds),   are treated simultaneously. 
Fixing topological numbers then corresponds to choosing a connected component of the total moduli space of solutions to (\ref{eqs}).

The (super)Yang-Mills theories (\ref{action}) on Morita equivalent noncommutative tori $T_{\theta}$, $T_{\hat \theta}$ 
are physically equivalent (see \cite{ASMorita}).
To any module $E$ over $T_{\theta}$ there corresponds a module $\hat E$ over $T_{\hat \theta}$ and to any connection 
$\nabla$ on $E$ there corresponds a connection $\hat \nabla$ and vice versa. The correspondence of connections 
respects gauge equivalence relation and maps a constant curvature connection into a constant curvature one. This 
means that moduli spaces of constant curvature connections are isomorphic for Morita equivalent tori.  
 A crucial fact that we are going to exploit analyzing representations of (\ref{eqs}) is that by use of Morita equivalence 
one can reduce the problem to the situation when $f_{ij}$ is a nondegenerate matrix (see Appendix).

\section{Constant curvature connections on noncommutative tori}
Let $A_{\theta}$ be a $d=2g$-dimensional noncommutative torus.
Let $E$ be a projective module over $A_{\theta}$ admitting a constant
curvature connection $\nabla_{i}$, $i=1,\dots , d$ 
\begin{equation} \label{f}
[\nabla_{j}, \nabla_{k} ] = 2\pi  if_{jk} \cdot {\bf 1} 
\end{equation}
where $f_{jk}$ is a constant nondegenerate antisymmetric matrix with real entries. 
(We assume that $\nabla_{i}$ are antihermitian operators.)
Up to normalizations  algebra (\ref{f}) is the Heisenberg algebra 
specified by  canonical commutation relations for $g$ degrees of freedom. 
According to the Stone-von Neumann theorem there is a unique irreducible representation  $\cal F$ of  algebra (\ref{f}).
We assume that $E$ can be decomposed into a direct sum of a finite number of irreducible components: 
$E\cong {\cal F}^{N}\cong {\cal F}\otimes {\bf C}^{N}$.

We fix the representation $\cal F$ as follows. First let us bring the matrix $f_{ij}$ to a canonical block-diagonal form 
\begin{equation} \label{cform}
(f_{ij})= \left( \begin{array}{cccc}
f_{1} {\bf \epsilon} & 0 & \dots & 0 \\
0 & f_{2} {\bf \epsilon} & \dots & 0 \\
0 & 0 & \ddots & 0 \\
0 & 0 & \dots  & f_{g}  {\bf \epsilon} 
\end{array} \right) 
\end{equation}
where 
\begin{equation} \label{epsilon}
{\bf \epsilon} = \left(
\begin{array}{cc} 0 & 1 \\ 
-1 & 0 \end{array} \right) 
\end{equation}
is a $2\times 2$ matrix 
and $f_{i}$ are positive numbers. 

Then we can define a representation space as $L_{2}({\bf R}^{g})$ 
and the operators $\nabla_{i}$  as 
\begin{equation} \label{nabla} 
\nabla_{j} = \sqrt{f_{(j+ 1)/2}}\partial_{j} \, , j - \mbox{odd} \, , \quad \nabla_{j} = i\sqrt{f_{j/2}}x_{j-1}\, , j - \mbox{even} 
\end{equation} 
where $\partial_{j}$, $x_{k}$, $j,k=1,\dots , g$  are derivative and multiplication by $x^{k}$ operators acting 
on smooth functions $f(x)\in L_{2}({\bf R}^{g})$.
An arbitrary representation of the torus generators $U_{i}$, $i=1, \dots, d$ has the form 
$
U_{i} = U_{i}^{st}\cdot u_{i} 
$  
where $U_{i}^{st}$ is some standard representation satisfying 
$$
[\nabla_{j}, U_{k}^{st}]= 2\pi i \delta_{jk} U_{k}^{st}
$$ 
and $u_{i}$ is an $N\times N$  unitary matrix. This form of representation of $U_{i}$ follows from the irreducibility 
of representation $\cal F$. 
A straightforward calculation shows that one can take $U_{i}^{st}$ to be   
\begin{equation}\label{Ust}
U_{k}^{st} = e^{-(f^{-1})^{kl}\nabla_{l}} \, .
\end{equation}
These operators satisfy
\begin{equation} \label{stcom}
U_{j}^{st}U_{k}^{st} = e^{-2\pi i (f^{-1})^{jk}} U_{k}^{st}U_{j}^{st} \, .
\end{equation}
Since $U_{i}= U_{i}^{st}\cdot u_{i}$ must give a representation of a noncommutative torus it follows from (\ref{stcom}) that 
so must do  the operators $u_{i}$. But the last ones are finite-dimensional matrices so they can only represent a 
noncommutative torus whose noncommutativity matrix has rational entries, i.e. $u_{i}$'s have to satisfy  
\begin{equation} \label{u}
u_{i}u_{j} = e^{2\pi i n^{ij}/N} u_{j}u_{i}
\end{equation}
where $N$ is a positive  integer and $n^{ij}$ is an integer valued antisymmetric matrix. 
Putting the formulas (\ref{stcom}) and (\ref{u}) together one finds that $U_{i}$'s give a representation of a noncommutative torus 
$T_{\theta}$ with 
\begin{equation} \label{theta}
\theta^{ij} = -(f^{-1})^{ij} + n^{ij}/N \, . 
\end{equation}

 It follows from the results obtained by M.~Rieffel (\cite{RieffelProj}) that 
for finite $N$ (i.e. when $E$ decomposes into a finite number of irreducible components) 
 the module $E$ endowed with $U_{i}=U^{st}_{i}\cdot u_{i}$ as above is a finitely generated projective module over $T_{\theta}$ with 
$\theta$ given in (\ref{theta}).  Conversely one can show that the finiteness of $N$ is required by the condition of 
$E$ to be finitely generated and projective. (See \cite{AstSchw} for a detailed discussion of modules admitting a 
constant curvature connection.)


{\it Topological numbers.}\\
Let us calculate here the topological numbers of the modules constructed above.
We assume here that the matrix $\theta_{ij}$ given in  (\ref{theta}) has irrational entries. Then a  
projective module $E$ is uniquely characterized by an integral element $\mu(E)$ of the even part of  Grassmann algebra 
$\Lambda^{even}({\bf R}^{d})$.  
In order to calculate $\mu(E)$ we can use  Elliot's formula
\begin{equation} \label{ell}
\mu(E) = exp\left( \frac{1}{2}\frac{\partial}{\partial \alpha^{i}} \theta^{ij}   \frac{\partial}{\partial \alpha^{j}} \right) ch(E) 
\end{equation}
where 
\begin{equation}\label{ch}
ch(E) = D\cdot  exp(\frac{1}{2\pi i} \alpha^{i} f_{ij} \alpha^{j}  ) 
\end{equation}
and $D$ is a nonnegative real number that plays a role of the dimension of a module in noncommutative geometry.  
From (\ref{ell}) applying a Fourier transform we obtain
\begin{eqnarray} \label{mu}
\mu(E) &=& D\cdot {\rm Pfaff}(f) \cdot \int d\beta \, 
exp( \frac{1}{2}\beta_{i} ((f^{-1})^{ij} + \theta^{ij})\beta_{j} + \alpha^{i}\beta_{i}) = \nonumber \\
&=&   D\cdot {\rm Pfaff}(f)  \cdot \int d\beta \, exp(\frac{1}{2}\beta_{i} n^{ij} \beta_{j} /N  + \alpha^{i}\beta_{i}) \, .
\end{eqnarray}
At this point it is convenient to assume that the matrix $n_{ij}$ is brought 
to a canonical block-diagonal 
form similar to (\ref{cform}) with integers $n_{i}$, $i=1,\dots, g$ on the diagonal  by means of an $SL(d, {\bf Z})$ transformation 
(this is always possible, see \cite{Igusa}). 
Then we can explicitly do the integration in (\ref{mu}) and obtain 
\begin{equation}\label{mu2}
\mu(E) = C\cdot \left(\frac{n_{1}}{N} + \alpha^{1}\alpha^{2}\right) \left(\frac{n_{2}}{N} + \alpha^{3}\alpha^{4}\right)\cdot \dots 
\cdot \left(\frac{n_{g}}{N} + \alpha^{2g-1}\alpha^{2g}\right)
\end{equation}
where $C=D\cdot {\rm Pfaff}(f) $ is a constant that can be determined by the requirement that $\mu(E)$ is an integral element 
of Grassmann algebra $\Lambda$ (i.e. each coefficient  is an integer). By looking at the term with a maximal number 
 of $\alpha$'s in (\ref{mu2}) we immediately realize that $C$ must be an integer.   We will prove below that $C=N$.

Let us introduce  numbers
$$
N_{i} = \frac{N}{g.c.d. (N, n_{i})} \, , \qquad \tilde N_{i} = \frac{n_{i}}{g.c.d. (N, n_{i})}
$$
so that for each $i=1, \dots , g$ the pair   $N_{i}$, $\tilde N_{i}$ is relatively prime.
Then we can rewrite (\ref{mu2}) as 
\begin{equation} \label{mu3} 
\mu(E) = \frac{C}{N_{1}N_{2}\cdot \dots \cdot N_{g}} \prod_{i=1}^{g} (\tilde N_{i} + N_{i}\alpha^{2i-1}\alpha^{2i}) \equiv 
\frac{C}{N_{1}N_{2}\cdot  \dots \cdot N_{g}}\mu_{0}(E) \, .
\end{equation}
For any integral element $\nu$ of Grassmann algebra $\Lambda$ let us introduce a number $g.c.d. (\nu)$ which is 
defined to be  the largest integer $k$ such that  $\nu = k\nu'$ where 
$\nu'$ is also integral.   
It is a simple task to prove by induction in $g$ that  $g.c.d. (\mu_{0}(E))=1$. 
Hence, $C$ must be an integer divisible by the product $N_{1}N_{2}\cdot \dots \cdot N_{g}$. Moreover 
 $C = g.c.d. (\mu(E) )N_{1}N_{2}\cdot \dots \cdot N_{g}$.  
It is known (for example see \cite{vanBaal})  that the dimension of  an irreducible representation of the algebra (\ref{u}) is equal to 
the product $N_{1}\cdot N_{2} \cdot \dots \cdot N_{g}$. 
Thus,  necessarily this product divides $N$, i.e. $N =  N_{1}\cdot N_{2} \cdot \dots \cdot N_{g}\cdot N_{0}$ 
where $N_{0}$ is an integer equal to the  number of  irreducible components in the  
representation ${\bf C}^{N}$ of the algebra (\ref{u}). 
Evidently $N_{0}$ divides $g.c.d.(\mu(E))$. We are going to show below that  $g.c.d.(\mu(E))$ cannot be bigger than $N_{0}$.
This will prove that $C=N$.

Let us look at some particular examples of formula (\ref{mu2}). 
 If the matrix $n^{ij}$ is nondegenerate then $\mu(E)$ is a quadratic exponent:
\begin{equation}
\mu(E) = p\cdot exp( \frac{1}{2}\alpha^{i}(n^{-1})_{ij} \alpha^{j} N) \, , \quad p=N_{0}\tilde N_{1}\cdot \dots \cdot \tilde N_{g} 
\end{equation} 
where $p=N\cdot {\rm Pfaff}(n/N)$ is written in a form where it is manifestly an integer.  
If $n^{ij}$ is degenerate then $\mu(E)$ is a so called generalized quadratic exponent (see \cite{AstSchw} and \cite{KS}, Appendix D). 
For example if $n^{ij}=0$ for all $i$ and $j$ then we obtain from (\ref{mu}) 
\begin{equation}
\mu(E) = N \alpha^{1}\alpha^{2}\cdot \dots \cdot \alpha^{d} \, .
\end{equation}

            
{\it Moduli space of constant curvature connections in terms of irreps of rational n.c. tori.}\\
In the previous subsection we showed that modules endowed with a constant curvature connection correspond to  representations of 
matrix algebra (\ref{u}). The residual gauge transformations preserving (\ref{nabla}) correspond to $N\times N$ 
unitary transformations acting on the ${\bf C}^{N}$ factor  of $E$. Thus, we see that the moduli space of 
 constant curvature connections on a module with fixed $(N, n^{ij})$ (or fixed $\mu(E)$, 
which is the same)  can be described as a space of inequivalent representations of the matrix algebra (\ref{u}).
 The center of algebra  (\ref{u}) is spanned by elements $u_{\bf k}$ with ${\bf k}\in D\cong {\bf Z}^{d}$ satisfying 
$k_{i}n^{ij}m_{j}/N \in {\bf Z}$ for any ${\bf m}\in D$. Such elements correspond to a sublattice of $D$ that we denote $D^{*}$ 
(the notation reveals the fact that $D^{*}$ can be viewed as a dual lattice to $D$). 
To describe the center in a more explicit way it is convenient to choose the basis we used above, in which the matrix  $n^{ij}$ 
is brought to a block-diagonal canonical form. 
In this basis  generators of the center can be chosen to be elements  $(u_{i})^{M_{i}}$ where we set 
 $M_{i} = N_{(i+1)/2}$, $i$-odd and  $M_{i} = N_{i/2}$, $i$-even.
Thus, in an irreducible 
representation $(u_{i})^{M_{i}}=\lambda_{i} \in {\bf C}^{\times}$ are constants of absolute value 1. 

Using the substitution
\begin{equation}\label{subst}
u_{i} \mapsto c_{i} u_{i}
\end{equation}
where $c_{i}$ are constants, $|c_{i}|=1$,  we obtain an irreducible representation with $\lambda_{i}'=\lambda_{i}...$ 
with values of center generators $\lambda_{i}$ into the one with $\lambda_{i}'=\lambda_{i}c_{i}^{M_{i}}$. 
 By means of this substitution one can transform any irrep into the one with $\lambda_{i} =1$. The last one corresponds to a representation of the 
algebra specified by relations (\ref{u}) along with the relations $(u_{i})^{M_{i}} =1$.
 This algebra has 
 unique irreducible representation of dimension  $N_{1}\cdot N_{2} \cdot \dots \cdot N_{g}$ (for example see \cite{vanBaal}). 
Therefore, the space of irreducible representations of algebra (\ref{u}) is described by means of $d$
complex numbers $\lambda_{i}$ with absolute value 1, i.e. is isomorphic to a (commutative) torus $\tilde T^{d}\cong {\bf R}^{d}/D^{*}$. 
We denote the corresponding 
irreps by $E_{\Lambda}$, $\Lambda=(\lambda_{1}, \dots , \lambda_{d})$.
In general for any noncommutative torus $T_{\theta}$ one can construct a group $L_{\theta}$ of automorphisms isomorphic to 
a commutative torus of the same dimension by means of (\ref{subst}). This torus acts naturally on the space of 
unitary representations of $T_{\theta}$. If $\theta$ is rational we obtain a transitive action of this automorphism 
group on the space of irreducible representations. In this case one can consider $L_{\theta}$ as a finite covering 
of $\tilde T^{d}$.

Let us assume now that the space   ${\bf C}^{N}$ is decomposed into irreducible representations of algebra (\ref{u})
\begin{equation} \label{dec}
{\bf C}^{N} = E_{\Lambda_{1}}\oplus \dots \oplus E_{\Lambda_{N_{0}}}\, .
\end{equation}
Note that in the picture we are working with,  gauge transformations are given by unitary linear operators acting on $E$ that 
commute with all $\nabla_{i}$'s that is by unitary $N\times N$ matrices.  
 The matrices representing central elements are diagonalized in the basis specified by  decomposition (\ref{dec}). 
There are residual gauge transformations corresponding to permutations of diagonal entries. Thus, we see that in  general 
 the moduli space is isomorphic to $(\tilde T^{d})^{N_{0}}/S_{N_{0}}$. 
As it was noted   in the previous subsection $N_{0}$ divides $g.c.d. (\mu(E))$.
On the other hand as we know from \cite{AstSchw}, \cite{KS} any module $E$ over a noncommutative torus $T_{\theta}$ 
admitting a constant curvature connection $\nabla_{i}$ 
can be represented as a direct sum of $k$ identical modules  $ E=E'\oplus \dots \oplus E' $ with 
$k=g.c.d.(\mu(E))$. This implies that the moduli space of constant curvature connections necessarily contains 
a subset isomorphic to $(\tilde T^{d})^{k}/S_{k}$. Thus, on dimensional grounds we conclude that $k=g.c.d.(\mu(E))=N_{0}$.

\section{General toroidal orbifolds}
In this section we will consider  how the considerations above can be generalized to the case of  toroidal orbifolds. 
We will work out in detail the particular cases of ${\bf Z}_{2}$ and ${\bf Z}_{4}$ orbifolds in the subsequent sections. 
Let $D\subset {\bf R}^{d}$ be a $d$-dimensional lattice  embedded in ${\bf R}^{d}$ and let $G$ be a finite 
group acting on   ${\bf R}^{d}$ by linear transformations mapping the lattice $D$ to itself. 
For an element $g\in G$ we will denote the corresponding representation matrix $R_{i}^{j}(g)$.
One can write down constraints describing compactification of M(atrix) theory on the orbifold $T^{d}/G$, 
where $T^{d}={\bf R}^{d}/D$: 
\begin{equation} \label{e1}
X_{j} + \delta_{ij}2\pi \cdot {\bf 1} = U_{i}^{-1}X_{j}U_{i} \, , 
\end{equation}
\begin{equation} \label{eq1'}
  X_{I} = U_{i}^{-1}X_{I}U_{i} \, \qquad \psi_{\alpha} = U_{i}^{-1}\psi_{\alpha}U_{i} \, ,
\end{equation}
\begin{equation} \label{e2}
R_{i}^{j}(g)X_{j} = W^{-1}(g)X_{i}W(g) \, , 
\end{equation}
\begin{equation} \label{eq2'}
 \Lambda_{\alpha \beta}(g)\psi_{\beta} = W^{-1}(g)\psi_{\alpha}W(g) \, , \quad X_{I} = W^{-1}(g) X_{I} W(g) \, . 
\end{equation}
Here  $i,j =1, \dots , d$ are indices for directions along the torus , $I= d+1,\dots , 9$ is an index corresponding to  the transverse 
directions, $\alpha$ is a spinor index; $\Lambda_{\alpha \beta}(g)$ is the matrix of spinor representation of $G$ obeying 
$\Lambda^{\dagger}(g)\Gamma^{i}\Lambda(g)=R_{ij}(g)\Gamma_{j}$; $ U_{i}$, $W(g)$ - unitary operators.  
One can check that the quantities $U_{i}U_{j}U_{i}^{-1}U_{j}^{-1}$ 
commute with all $X_{i}$, $X_{I}$, and $\psi_{\alpha}$. It is natural to set them to be proportional to the identity operator. 
This gives us defining relations of  a noncommutative torus
$$
U_{j}U_{k} = e^{2\pi i\theta_{jk}}U_{k}U_{j} \, .  
$$
It is convenient to work with linear generators $U_{\bf n}$ that can be expressed in terms of products of $U_{i}$.  
One can further check that expressions $W(gh)W^{-1}(g)W^{-1}(h)$ and $W^{-1}(g)U_{\bf n}W(g)U^{-1}_{R^{-1}(g){\bf n}}$
also commute with all  fields $X_{i}$, $X_{I}$, $\psi_{\alpha}$. We assume that these expressions are proportional 
to the identity operator. This leads us to the following relations  
\begin{eqnarray}
&&W(g)W(h) = W(gh) e^{i\phi(g,h)} \, , \nonumber\\ 
&& W^{-1}(g)U_{\bf n}W(g) = U_{R^{-1}(g){\bf n}} e^{i \chi({\bf n},g)}
\end{eqnarray}
where $\phi(g,h)$, $\chi({\bf n},g)$ are constants.
The first equation means that operators $W(g)$ furnish a projective representation of $G$. 
It follows from these equations that the matrix $\theta$ is  invariant under the group action $R(g)$.  
Below we will confine ourselves to the case  of vanishing  cocycles $\phi$ and $\chi$. We refer the reader to 
papers \cite{HoWu}, \cite{Doug_discr} for a discussion of cases when  cocycle $\phi$ does not vanish. 
(Note that for cyclic groups both cocycles are always trivial. This means that they can be absorbed into 
redefinitions of generators.)

One can define an algebra of functions on a  noncommutative orbifold as an algebra generated by 
the operators $U_{\bf n}$ and  $W(g)$ satisfying  (\ref{nct}) and 
\begin{equation} \label{WU}
 W^{-1}(g)U_{\bf n}W(g) = U_{R^{-1}(g){\bf n}} \, , 
\end{equation}
\begin{equation}
  W(g)W(h) = W(gh) \, .
\end{equation}
These equations  mean that the algebra at hand is a crossed product 
$T_{\theta}\rtimes_{R} G$. Again we remark here that allowing central extensions gives a more general case of 
twisted crossed products. In this paper we will concentrate on the untwisted case.

The algebra $T_{\theta}\rtimes_{R} G$ can be equipped by an involution $*$ by setting 
$U_{\bf n}^{*} = U_{-{\bf n}}$, $W^{*}(g) = W(g)$. This makes it possible to embed these algebras 
into a general theory of $C^{*}$ algebras. 
A projective module over  an orbifold can be considered as a projective module $E$
over $T_{\theta}$ equipped with operators $W(g)$, $g\in G$ satisfying (\ref{WU}).
The equations (\ref{e1}), ({\ref{e2})  mean that $X_{i}$ specifies
a $G$- equivariant connection  on $E$, i.e. $\nabla_{i}$ is  a $T_{\theta}$ connection  satisfying 
\begin{equation}\label{eqcon}
R_{i}^{j}(g)\nabla_{j} = W^{-1}(g)\nabla_{i}W(g) \, .
\end{equation}
The fields $X_{I}$ are endomorphisms of $E$, commuting both with $U_{\bf n}$ and $W(g)$ and the spinor fields 
 $\psi_{\alpha}$ can be called equivariant spinors.

Let us comment here  on the supersymmetry of these compactifications. 
The surviving supersymmetry transformations are transformations (\ref{susy}) corresponding 
to invariant spinors $\epsilon$, $\tilde \epsilon$, i.e. the ones satisfying 
$\Lambda(g)\epsilon = \epsilon$. For $d=4$, $6$  this equation has a nontrivial solution 
provided the representation $R(g)$  lies  within  an $SU(2)$, $SU(4)$ subgroup respectively.
 The possible finite groups $G$ that can 
be embedded in this way are well known. Those include the examples of ${\bf Z}_{2}$ and 
${\bf Z}_{4}$ four-dimensional orbifolds to be considered below. In those cases when the 
supersymmetry is not broken completely  equivariant connections of constant curvature 
correspond to configurations preserving half of the unbroken supersymmetries.

Now we restrict ourselves to modules admitting a constant curvature equivariant connection (\ref{f}). 
 All the steps of the analysis made above for the case of tori leading to the decomposition (\ref{dec}) can 
be repeated in a straightforward way. We should add now to 
that analysis operators $W(g)$. 
 Equation (\ref{eqcon}) implies that the curvature tensor $f_{ij}$ is invariant under the action of $G$. 
As above we fix a representation of the Heisenberg algebra (\ref{f}) in the form (\ref{nabla}).
Then a connection $\nabla^{g}_{i}= R_{i}^{j}(g)\nabla_{j}$ gives a representation of the same Heisenberg 
algebra (\ref{f}). By uniqueness of irreducible representation $\cal F$  there exists a set of operators 
$W^{st}(g): {\cal F} \to {\cal F}$ satisfying (\ref{eqcon}).   
It follows from the definition  (\ref{Ust}) of $U^{st}_{j}$  that $W^{st}(g)$ and $U^{st}_{j}$ 
commute as  in (\ref{WU}). This implies that a general set of operators $W(g):E\to E$ has a form 
\begin{equation}
W(g) = W^{st}(g)\otimes w(g) 
\end{equation}
where $w(g)$ are $N\times N$ matrices satisfying 
\begin{equation} \label{wu}
w^{-1}(g) u_{\bf n} w (g) =  u_{R^{-1}(g){\bf n}} \, .  
\end{equation}
Here $u_{\bf n}$ are (linear) generators  of the rational torus (\ref{u}). 
 In general  relation (\ref{eqcon}) only implies that  
the operators $W^{st}(g)$  form a projective representation of $G$. Then,  the commutation relations 
for $w(g)$ are twisted by an opposite cocycle. 
The problem of describing moduli space of equivariant constant curvature connections 
now boils down to the study of irreducible representations of the matrix algebra generated by $u_{i}$, $w(g)$.

As it was explained above an irreducible representation of a rational torus is labeled by a point of a  
commutative torus $\tilde T^{d}= {\bf R}^{d}/D^{*}$. The lattice $D^{*}$ is a sublattice of $D$ 
which is preserved by $G$. Therefore $G$ acts on the torus $\tilde T^{d}$.
 We denote this action by $R^{*}(g)$.
It follows from (\ref{wu})  that an irreducible representation $E_{\Lambda}$ from decomposition (\ref{dec}) is mapped by  
$w(g)$ into $E_{R^{*}(g) \Lambda}$. Therefore, the set of $\Lambda_{i}$ in the decomposition (\ref{dec}) has to be invariant 
under the action $R^{*}(g)$. The irreducible representations of the algebra generated by $u_{i}$, $w(g)$ can be grouped 
according to the  orbits of $G$ action on $\tilde T^{d}$. For a generic point one has an orbit of length $|G|$. 
There are also exceptional points that include
fixed points and points whose orbits length is a nontrivial divisor of $|G|$. 
If the exceptional points are isolated the corresponding
 representations can be interpreted in terms of branes that are 
stuck at the exceptional points. 
Below we will consider in detail the structure of the aforementioned representations for the orbifold 
groups ${\bf Z}_{2}$ and ${\bf Z}_{4}$.

\section{Noncommutative $\bf Z_{2}$ orbifolds}
As in the case of tori we start by fixing the representation of connection (\ref{nabla}). In addition to representation 
of the torus generators (\ref{Ust}), (\ref{u}) now we need to represent the $\bf Z_{2}$ generator $W$. 
As in the case of $U_{i}$'s a generic representation of $W$ has the form $W=W^{st}w$ where $W^{st}$ is some standard representation 
satisfying 
$$
W^{st}\nabla_{i}W^{st} = - \nabla_{i} \, , \quad W^{st}U_{i}^{st}W^{st} = (U_{i}^{st})^{-1} \, , \quad (W^{st})^{2} = 1 
$$
and $w$ is a $N\times N$ matrix that obeys the relations 
\begin{equation}\label{w}
wu_{i}w=u_{i}^{-1} \, , \qquad w^{2} = 1 \, .
\end{equation}
It is easy to check that one can take 
$$
W^{st} : f_{i}(x) \mapsto f_{i}(-x) 
$$
where $f_{i}(x) \in E\cong L_{2}({\bf R}^{d})\otimes {\bf C}^{N}$ ($i=1, \dots , N$). 
As before consider  the decomposition of  ${\bf C}^{N}$ into irreducible representations of the algebra (\ref{u}) given by (\ref{dec}).  
 The operator $w$ maps $ E_{\Lambda} \to E_{\Lambda^{-1}}$ 
where $\Lambda^{-1} \equiv (\lambda_{1}^{-1}, \dots , \lambda_{d}^{-1})$. Thus, the space ${\bf C}^{N}$ carries a representation 
of the algebra specified by (\ref{u}) and (\ref{w}) only if the set of $\{ \Lambda_{i}\, , i=1, \dots , k\}$ in (\ref{dec}) is 
invariant under the inversion $\Lambda_{i} \mapsto \Lambda_{i}^{-1}$. We see that  decomposition (\ref{dec}) can contain
summands in the form  of couples $\tilde E_{\Lambda}\equiv E_{\Lambda}\oplus E_{\Lambda^{-1}}$  and  possible  exceptional representations 
$E_{\Lambda_{\epsilon}}$ with $\Lambda_{\epsilon} = (\epsilon_{1}, \dots , \epsilon_{d})$, $\epsilon_{i} = \pm 1$, i.e. we have 
\begin{equation} \label{dec2}
{\bf C}^{N} = \left( \tilde E_{\Lambda_{1}} \oplus \dots \oplus \tilde E_{\Lambda_{r}}\right) \bigoplus_{\epsilon} E_{\Lambda_{\epsilon}}^{\tau (\epsilon )}
\end{equation} 
where $\tau (\epsilon )$ are nonnegative integers. 
This decomposition can be chosen in such a way that the  matrix $w$ has a  block diagonal form 
with $r$  blocks $w_{i}: \tilde E_{\Lambda_{i}}\to \tilde E_{\Lambda_{i}}$ of the form  
$$
 w_{i} = \left(
\begin{array}{cc} 0 & \tilde w_{i} \\ 
\tilde w_{i}^{-1} & 0 \end{array} \right) 
$$
where $\tilde w_{i} : E_{\Lambda_{i}} \to  E_{\Lambda_{i}^{-1}}$ and a number of blocks $w_{\epsilon} : E_{\Lambda_{\epsilon}} \to E_{\Lambda_{\epsilon}}$.

Let us first look at the components specified by pairs  $(E_{\Lambda_{\epsilon}}, w_{\epsilon})$. It follows from the irreducibility 
of the representation  $E_{\Lambda_{\epsilon}}$ and the fact that $w_{\epsilon}^{2}=1$ that $w_{\epsilon}$ is defined up to an 
overall sign. Let $w^{(0)}_{\epsilon}$ be  a standard choice of $w_{\epsilon}$. Denote by  $F^{\pm}_{\epsilon}$ the representation 
of the algebra (\ref{u}), (\ref{w})  
specified by  $(E_{\Lambda_{\epsilon}}, \pm w_{\epsilon}^{(0)})$.

As for the blocks $( \tilde E_{\Lambda_{i}}, w_{i})$ we first note that due to the irreducibility of $E_{\Lambda_{i}}$ and 
$ E_{\Lambda_{i}^{-1}}$ the operator $\tilde w_{i}$  is defined up to a  constant factor. 
This constant factor can be gauged away by conjugating $w_{i}$ with a suitable rescaling transformation 
$$
 \left(
\begin{array}{cc} {\bf 1}\cdot \mu_{1} & 0 \\ 
 0 &  {\bf 1}\cdot \mu_{2} \end{array} \right) 
$$
where $\mu_{1}$, $\mu_{2}$ are complex numbers of absolute value 1 (these rescalings are allowed gauge transformations).
Thus,  
we get a single representation of  the algebra specified by $(u_{i}, w)$  that we denote $F_{\Lambda_{i} }$. 
This representation is irreducible except the cases when  $\Lambda_{i} = \Lambda_{\epsilon}$ for some $\epsilon$. 
Then  
we have $\tilde w_{i} = \pm w^{(0)}_{\epsilon}$ and  one can readily check that 
$F_{\Lambda_{\epsilon}} \cong F_{\epsilon}^{+} \oplus F_{\epsilon}^{-}$. 
This permits one to decompose ${\bf C}^{N}$ into the components 
\begin{equation} \label{dec'}
{\bf C}^{N} = \left( F_{\Lambda_{1}}\oplus \dots \oplus F_{\Lambda_{r} } \right) 
\bigoplus_{\epsilon} (F_{\epsilon}^{\eta_{\epsilon}})^{\tau(\epsilon)} 
\end{equation}
where $\eta_{\epsilon}=\pm $, $\tau(\epsilon)$ are nonnegative integers specifying the multiplicities with which 
the corresponding representations enter the decomposition. 
Note that the set of numbers $\Lambda_{i}$, $r$, $\tau(\epsilon)$ is  uniquely determined by the given 
$B_{\theta}= A_{\theta}\rtimes {\bf Z}_{2}$ module $E$ and an equivariant constant curvature connection $\nabla_{i}$ on it. 
The residual gauge transformations preserving the decomposition (\ref{dec'}) can be represented as  compositions 
of transpositions  acting inside each  $ F_{\Lambda_{i}}$ block and sending $\Lambda_{i} \to \Lambda_{i}^{-1}$ 
and permutations of different $ F_{\Lambda_{i} }$ blocks.  Thus, the moduli space of equivariant constant curvature  connections 
 is isomorphic to  $(T^{d}/{\bf Z_{2}})^{r}/S_{r}$ where $r$ is some integer. Using the relation 
$$
 g.c.d. (\mu(E)) = 2r + \sum_{\epsilon} \tau(\epsilon)
$$
one can find an estimate from above on $r$: $r\le \Bigl[ \frac{g.c.d.(\mu)}{2} \Bigr]$.

Let us comment here briefly on the invariance of the above results under  Morita equivalence. It follows from the definition  that the mapping 
of modules and connections induced by (complete) Morita equivalence relation preserves the gauge equivalence relation and 
maps constant curvature connections into constant curvature connections. Hence the moduli spaces over the modules related by Morita equivalence are 
isomorphic.
In particular the  dimension $r$ of the moduli space  is preserved under Morita equivalences. As the number $g.c.d. (\mu(E))$ 
is also preserved we see that 
$\sum_{\epsilon} \tau(\epsilon) = g.c.d. (\mu(E)) - 2r$
is preserved under Morita equivalence. In fact the set of pairs  $(\tau(\epsilon), \eta_{\epsilon})$ can only get 
permuted under  Morita equivalence transformations. One can show that for modules admitting a constant curvature connections 
the set of parameters  $\{(\tau(\epsilon), \eta_{\epsilon})\}$ is in a one-to-one correspondence with additional 
topological numbers characterizing the module (see \cite{z2}).

\section{${\bf Z}_{4}$ orbifolds}
In this section we will consider the case of toroidal ${\bf Z}_{4}$ orbifolds. For definiteness let us 
concentrate on the four-dimensional case. It is not hard to extend the results we  obtain to other 
even-dimensional  ${\bf Z}_{4}$ toroidal orbifolds. For $d=4$ without loss of generality one can define a  ${\bf Z}_{4}$ action on ${\bf R}^{4}$ by
$\rho: (x,y) \mapsto (-y, x)$. Here we assume that ${\bf R}^{4}$ has a product structure ${\bf R}^{2}\times  {\bf R}^{2}$ and 
    $x$, $y$  belong   to the first and second ${\bf R}^{2}$ factor respectively. This action preserves the orthogonal 
lattice $D\cong {\bf Z}^{4}\subset {\bf R}^{4}$ and thus descends to an action on the torus $ {\bf R}^{4}/D$. 
We can consider a noncommutative four-torus $T_{\theta}$ constructed by means of lattice $D$ and an antisymmetric 
two-form $\theta_{ij}$ that is assumed to be decomposed into a $2\times 2$ block form relative to the  ${\bf R}^{2}\times  {\bf R}^{2}$ 
product   structure. One can easily see that such a form is invariant with respect to the above defined  ${\bf Z}_{4}$ action.  
Thus, we can consider a noncommutative toroidal orbifold $T_{\theta}\rtimes_{\rho}{\bf Z}_{4}$.

Let $E$ be a projective module over this orbifold and 
  let $Y$ denotes a representation of the ${\bf Z}_{4}$ generator.  
Applying  the general construction of section ... we arrive at the decomposition (\ref{dec}) and a representation of 
a matrix algebra generated by $u_{i}$ and $y$, satisfying (\ref{u}) and 
\begin{equation} \label{y}
y^{4} = 1 \, , \qquad y^{-1}u_{\bf n}y = u_{\rho^{-1}{\bf n}} \, . 
\end{equation}
The total representation space ${\bf C}^{N}$ splits into irreducible representations of this matrix algebra. 
To describe those we first classify  the orbits of the ${\bf Z}_{4}$  action $\rho^{*}$ on the dual torus $\tilde T^{4}\cong {\bf R}^{d}/D^{*}$. 
For a generic point of  $\tilde T^{4}$ the orbit consists of 4 distinct points. 
There are 16 exceptional points. Those include the four fixed points: $(0,0;0,0)$, $(0, 1/2;0, 1/2)$, $(1/2, 0; 1/2; 0)$, and $(1/2,1/2,1/2,1/2)$,  and 
 12 points whose orbit is of length 2. The last ones  have coordinates $0$ or $1/2$ and they complete the ${\bf Z}_{4}$ 
action fixed points written above to the whole set of 16 ${\bf Z}_{2}$ action fixed points (the  ${\bf Z}_{2}$ action is specified  by $\rho^{2}$).  
Hence, we have 4 orbits of order 1 and 6 orbits of order 2. Denote the fixed points $\Lambda_{\epsilon}$, $\epsilon=1, \dots , 4$ 
and the pairs of points whose orbits are of order 2 by $(\Lambda_{\nu}', \Lambda_{\nu}'')$, $\nu=1, \dots , 6$.

Let us first consider the representations corresponding to generic points. Denote 
$$
\tilde E_{\Lambda}=  E_{\Lambda}\oplus E_{\rho^{*}\Lambda}\oplus  E_{(\rho^{*})^{2}\Lambda}\oplus  E_{(\rho^{*})^{3}\Lambda}\, . 
$$  
Relative to this decomposition the generator $y$ can be written in a block form as 
$$
 y= \left(
\begin{array}{cccc}  
0& y_{12}&0&0 \\ 
 0 & 0& y_{23} & 0 \\
0&0&0 &y_{34} \\
y_{41}&0&0&0 
\end{array} \right) 
$$
where $y_{ij}:  E_{(\rho^{*})^{i-1}\Lambda}  \to E_{(\rho^{*})^{i}\Lambda}$. It follows from irreducibility 
of representations $E_{\Lambda}$ that the blocks $y_{ij}$ are determined uniquely up to  constant factors. 
 The last ones can be gauged away by gauge transformations that multiply each of the $E_{(\rho^{*})^{n}\Lambda}$
components of $\tilde E_{\Lambda}$ by a constant. Thus, we obtain a representation  $F_{\Lambda}$ of the matrix algebra 
(\ref{u}), (\ref{y})  (we hope that using the same notation as the one used in 
the previous section when studying the ${\bf Z}_{2}$ case will not cause any confusion).

Next let us look at representations labeled by pairs of exceptional points $(\Lambda_{\nu}', \Lambda_{\nu}'')$. 
The representation of rational torus is $\hat E_{\Lambda_{\nu}} = E_{\Lambda_{\nu}'}\oplus E_{\Lambda_{\nu}''}$. 
A general form of the  generator $y$ is 
$$
y= \left( 
\begin{array}{cc}
0 & \mu_{1} \cdot y_{1} \\
\mu_{2} \cdot y_{2} & 0 
\end{array}
\right)
$$  
where $y_{1}:  E_{\Lambda_{\nu}''} \to  E_{\Lambda_{\nu}'}$, $y_{2}: E_{\Lambda_{\nu}'}\to  E_{\Lambda_{\nu}''}$ are fixed 
and $\mu_{1}$, $\mu_{2}$ are constants satisfying $(\mu_{1}\mu_{2})^{2}=1$. Using gauge transformations 
one can bring $y$ to one of the two forms specified by $\mu_{1}=\mu_{2}=1$ and $\mu_{1}=-\mu_{2}=1$. 
Therefore, we have two inequivalent representations of the algebra (\ref{u}), (\ref{y}) denoted 
$G^{\pm}_{\nu}$ with $\pm$ standing for the sign of the product $\mu_{1} \mu_{2}$.

Finally, consider the fixed points  $\Lambda_{\epsilon}$. 
Since  each $E_{\Lambda_{\epsilon}}$ is an irreducible representation of a rational torus the operator $y$ acting on it 
is  defined uniquely  up to a multiplication by a 4-th root of unity $\xi_{k}=exp(2\pi ik/4)$, $k=0,1,2,3$ . 
Thus, for every fixed point we have 4 different representations $F_{\epsilon}^{k}$,  $k=0,\dots, 3$.

The generic representation  $F_{\Lambda}$ is irreducible unless $\Lambda$ hits one of the exceptional points. 
If it hits a fixed point the representation splits as $F_{\Lambda_{\epsilon}} \cong \oplus_{k=0}^{3} F_{\epsilon}^{k}$. 
In this case the representation of ${\bf Z}_{4}$ is a tensor product of a regular representation with a representation 
that acts on a single copy of $E_{\Lambda_{\epsilon}}$. If $\Lambda$ coincides with one of $\Lambda_{\nu}'$ or 
$\Lambda_{\nu}''$ the corresponding representation splits as 
$F_{\Lambda_{\nu}'} \cong F_{\Lambda_{\nu}''} \cong G^{+}_{\nu} \oplus G^{-}_{\nu}$. 
Using this equivalences we can decompose a general representation as 
\begin{equation} \label{z4dec}
{\bf C}^{N} = (F_{\Lambda_{1}}\oplus \dots \oplus F_{\Lambda_{r}}) \bigoplus_{\nu} (G^{\zeta_{\nu}}_{\nu})^{s(\nu)} 
\bigoplus_{\epsilon} \bigl( (F_{\epsilon}^{\eta_{\epsilon}^{1}})^{\tau^{1}(\epsilon)}\oplus 
 (F_{\epsilon}^{\eta_{\epsilon}^{2}})^{\tau^{2}(\epsilon)}\oplus(F_{\epsilon}^{\eta_{\epsilon}^{3}})^{\tau^{3}(\epsilon)} \bigr)
\end{equation}
where $\zeta_{\nu}=\pm$; $\eta_{\epsilon}^{i}$, $i=1,2,3$ are a triple of distinct integers from 0 to 4; 
$\tau^{i}(\epsilon)$ and $s(\nu)$ are nonnegative integers standing for the  multiplicities of modules.  
The numbers $r$, $\tau^{i}(\epsilon)$,  $s(\nu)$, $\zeta_{\nu}=\pm$, $\eta_{\epsilon}^{i}$ are uniquely determined by 
a given module. In is straightforward to generalize considerations of the previous section to show that  the moduli 
space  of equivariant constant curvature connections is isomorphic to $(\tilde T^{d}/{\bf Z}_{4})^{r}/S_{r}$ where 
$r$ is some integer such that $r\le \Bigl[ \frac{g.c.d.(\mu)}{4} \Bigr]$. The last restriction follows from the relation 
$$
g.c.d. (\mu) = 4r + 2(\sum_{\nu} s(\nu) ) + \sum_{\epsilon}\sum_{i=1}^{3} \tau^{i}(\epsilon) \, . 
$$

\section{ Coulomb branches of the moduli space}

Once we described moduli space of constant curvature connections it is not hard 
to add scalar fields to the discussion. Let us first consider the case of tori. The equations for scalars that we have are 
\begin{equation} \label{X}
[\nabla_{i}, X_{I}]= 0 \, , \qquad [X_{I}, X_{J}]=0 \, , \qquad [U_{i}, X_{I}]=0 \, .
\end{equation}
Here $I=1, \dots, 9-d$. 
For the fixed representation of $\nabla_{i}$ (\ref{nabla}) the first equation in (\ref{X}) implies that
 $X_{I}$ are constant $N\times N$ matrices. 
It follows then from the last two equations that the matrices $X_{I}$, $u_{i}$ can be simultaneously brought 
to the form when $X_{I}$'s are diagonal and $u_{i}$'s are block diagonal corresponding to the decomposition 
(\ref{dec}). Moreover the $X_{I}$'s are constant on each irreducible component $E_{\Lambda_{i}}$. Quotienting 
over $S_{N_{0}}$ residual gauge transformations gives us Coulomb branches isomorphic to $({\bf R}^{9-d})^{N_{0}}/S_{N_{0}}$ that 
matches with a dual (Morita equivalent) description of this system as a system
of $N_{0} = g.c.d.(\mu(E))$ D0-branes.

For the case of a ${\bf Z}_{2}$ orbifold in addition to equations (\ref{X}) we have the condition 
\begin{equation}
[W, X_{I}] = 0 \, .
\end{equation}
This implies that the matrices $X_{I}$ commute with the matrix $w$. Thus, the matrices  $u_{i}$, $w$, and $X_{I}$'s can be 
simultaneously brought to the form that corresponds to the decomposition (\ref{dec2}) and $X_{I}$ are in a block diagonal 
form with $r$ blocks $x_{I}^{(i)}: F_{\Lambda_{i}, \eta_{i}} \to  F_{\Lambda_{i}, \eta_{i}}$, $i=1, \dots, r$ and 
blocks of the form $x_{I}^{\epsilon}: F_{\epsilon} \to  F_{\epsilon}$. The last ones are necessarily constants due to 
irreducibility of $E_{\Lambda_{i}}$. If none of the points $\Lambda_{i}$ coincides with one of the fixed points $\Lambda_{\epsilon}$ 
we obtain  Coulomb branches of the form 
$$
({\bf R}^{(9-d)})^{r}/S_{r}\times 
\prod_{\epsilon}(({\bf R}^{(9-d)})^{\tau(\epsilon )}/S_{\tau(\epsilon )} )\, .
$$ If $\Lambda_{i_{k}} = \Lambda_{\epsilon}$ for some 
subset of indices $i_{k}$, $k=1,\dots , p$  
and for some  $\epsilon$, each representation $F_{\lambda_{i_{k}}, \eta}$ breaks into a direct sum of two representations $F_{\epsilon}$ 
and $F_{\epsilon}^{-}$ and instead of the factor 
$({\bf R}^{(9-d)})^{r}/S_{r}\times 
(({\bf R}^{(9-d)})^{\tau(\epsilon )}/S_{\tau(\epsilon )} )$
we get a factor of 
\begin{equation} \label{split}
({\bf R}^{(9-d)})^{r-p}/S_{r-p}\times ({\bf R}^{(9-d)})^{p}/S_{p}\times ({\bf R}^{(9-d)})^{\tau(\epsilon )+ p}/S_{\tau(\epsilon )+p}  \, .
\end{equation} 
This picture has an interpretation suggested in \cite{RamgWald} 
in terms of splitting of a D0 particle into a membrane-antimembrane pair that occurs once the D0 particle hits a  fixed point. 
In terms of this interpretation (\ref{split}) corresponds to $p$ membranes (or antimembranes depending on the value of $\eta_{\epsilon}$) 
and $p+ \tau(\epsilon )$ antimembranes (membranes) sitting at the fixed point $\Lambda_{\epsilon}$.

Now let us look at the ${\bf Z}_{4}$ case. Again we have a set of $9-d$ $N\times N$ matrices $X_{I}$. These matrices 
commute between themselves and with  the  matrix $y$. 
This leads us to a block decomposition of each of the $X_{I}$ relative to the decomposition (\ref{z4dec}) in which $X_{I}$ is 
constant on every irreducible representation of the algebra generated by $u_{i}$, $y$. 
Thus, for a generic point in the moduli space of constant curvature connection the Coulomb branch is 
$$
({\bf R}^{(9-d)})^{r}/S_{r}\times \prod_{\nu}({\bf R}^{(9-d)})^{s(\nu)}/S_{s(\nu)}
\prod_{\epsilon}\prod_{i=1}^{3}({\bf R}^{(9-d)})^{\tau^{i}(\epsilon)}/S_{\tau^{i}(\epsilon)} \, . 
$$  
If one of $\Lambda_{i}$ in (\ref{z4dec}) coincides with one of exceptional points $\Lambda_{\nu}'$, $\Lambda_{\nu}''$ or 
$\Lambda_{\epsilon}$ representations $F_{\Lambda_{i}}$ split accordingly as 
$F_{\Lambda_{\nu}'} \cong F_{\Lambda_{\nu}''} \cong G^{+}_{\nu} \oplus G^{-}_{\nu}$, or  
$F_{\Lambda_{\epsilon}} \cong \oplus_{k=0}^{3} F_{\epsilon}^{k}$. In that case the block decomposition of $X_{I}$ 
is  different and the answer is best described using  brane terminology. 
In general the moduli space coincides with the one describing  a system consisting of a number of free D0 particles, 
four different types of membranes that are stuck at fixed points and two types of couples of membranes sitting at points 
$(\Lambda_{\nu}', \Lambda_{\nu}'')$. It seems to be natural to identify different types of membranes  with a 
discrete $B$-field flux carried by them.  \\
\begin{center} {\bf Acknowledgements} 
\end{center}
We are grateful to J.~Blum, M.~Rieffel  and C.~Schweigert for stimulating discussions. 
\appendix
\section*{Appendix. Morita equivalence of toroidal orbifolds}

In this appendix we will show how Morita equivalences known for noncommutative tori can be extended to noncommutative 
toroidal orbifolds. We will give a general construction for arbitrary orbifold group $G$ and then specialize to the 
${\bf Z}_{2}$ and ${\bf Z}_{4}$ cases.

We start with a  reminding of some basic definitions concerning Morita equivalence (see \cite{ASMorita} for details).
Let $A$ and $\hat A$ be two involutive associative algebras. 
A $(A,\hat A)$-bimodule $P$ is said to  establish a   Morita equivalence between $A$ and 
$\hat A$. This means that the projective bimodule $P$ obeys the conditions 
\begin{equation} \label{P}
\bar P\otimes_{A}P\cong \hat A \, , \quad P\otimes_{\hat A}\bar P \cong A 
\end{equation}
where $\bar P$ is a $(\hat A, A)$-bimodule  that is  complex conjugated to $P$ that means that 
$\bar P$ consists of elements of $P$ and multiplications are defined as $\hat a (e) := (e) \hat a^{*}$, 
$ (e)a := a^{*} (e)$ where multiplications on the right hand sides are those defined for bimodule $P$.   
The algebras $A$ and $\hat A$ are said to be Morita equivalent if such $P$ satisfying  (\ref{P}) exists
The bimodule $P$ determines a one to one correspondence between $A$-modules and 
$\hat A$-modules by the rule 
$$
\hat E=E\otimes_{A}P \, , \quad E=\hat E\otimes_{\hat A}\bar P \, .
$$ 

For the case of noncommutative tori one can define a notion of complete Morita equivalence (\cite{ASMorita})
that allows one to transport connections between modules $E$ and $\hat E$. 
Let us remind here the basic  definitions. Let $\delta_{j}$, $j=1,\dots , d$ be a set of derivations of  $A_{\theta}$ 
specified by their action on generators  
$$
\delta_{j}U_{\bf n} = 2\pi i n_{j}U_{\bf n} \, .
$$ 
A connection on a projective module $E$ over $A_{\theta}$ can be defined as a set of operators 
$\nabla_{i}: E\to E$ satisfying a Leibniz rule 
$$
\nabla_{i}(ea) = (\nabla_{i}e)a + e(\delta_{i}a)
$$
for any $e\in E$ and any $a\in T_{\theta}$. 
We say that $(A_{\theta},A_{\hat \theta})$ Morita equivalence bimodule $P$ establishes a complete Morita equivalence if it
is endowed with operators $\nabla^{P}_{i}$ that determine a constant curvature connection 
simultaneously with respect to $A_{\theta}$ and $A_{\hat \theta}$, i.e. satisfy  

\begin{eqnarray}\label{biconnect}
&&\nabla^{P}_{i}(ae)=a\nabla^{P}_{i}e + (\delta_{i}a)e \, , \nonumber \\
&&\nabla^{P}_{i}(e\hat a)=(\nabla^{P}_{i}e)\hat a + e\hat \delta_{i} \hat a \, , \nonumber \\
&&[\nabla^{P}_{i},\nabla^{P}_{j}]=\sigma_{ij}\cdot {\bf 1} \, .
\end{eqnarray} 
Here  $\delta_{i}$ and $\hat \delta_{j}$ are standard derivations on $A_{\theta}$ and $A_{\hat \theta}$ respectively. 
Sometimes for brevity we will omit the word Morita in the term (complete) Morita equivalence bimodule.
If $P$ is a complete  $(A_{\theta},A_{\hat \theta})$ equivalence  bimodule then there exists a correspondence 
between connections on $E$ and connections on $\hat E$. An operator 
$\nabla_{i}\otimes 1 + 1\otimes \nabla_{i}^{P}$ on $E\otimes_{\rm C}P$ descends to a 
connection $\hat \nabla_{\alpha}$ on $\hat E = E\otimes_{A_{\theta}}P$. The curvatures of 
$\hat \nabla_{i}$ and $\nabla_{i}$ are connected by the formula 
$F_{ij}^{\hat \nabla}=\hat F_{ij}^{\nabla} + {\bf 1}\sigma_{ij}$.

Given a  $(A_{\theta},A_{\hat \theta})$ equivalence bimodule $P$ there is a possibility of extension of  the equivalence relation 
that it defines to 
modules over orbifolds $A_{\theta}\rtimes_{R}G$, $A_{\hat \theta}\rtimes_{R}G$. 
We will first describe a general construction and then discuss for which bimodules $P$ it exists.

 Assume that $P$ is equipped with a set of operators $W^{P}(g)$, $g\in G$ satisfying 
\begin{eqnarray} \label{WP}
&&W^{P}(g)W^{P}(h) = W(hg) \, , \nonumber \\
&& (W^{P}(g))^{-1}U_{\bf n} W^{P}(g) = U_{R{\bf n}} \, , \qquad    (W^{P}(g))^{-1}\hat U_{\bf n} W^{P}(g) = \hat U_{R{\bf n}}
\end{eqnarray}  
where $U_{\bf n}$ and $\hat U_{\bf n}$ stand for actions of $A_{\theta}$ and  $A_{\hat \theta}$ generators respectively. 
The first equation in (\ref{WP}) means that operators $W^{P}(g)$ give a right action of the group $G$ on $E$ that is 
just a choice of conventions. If $F$ is a right module over  $A_{\theta}\rtimes_{R}G$ specified by a $A_{\theta}$-module 
$E$ and  operators $W(g)$ acting on it one defines a  right $A_{\hat \theta}\rtimes_{R}G$ module $\hat F$ as 
a $A_{\hat \theta}$-module $\hat E = E\times_{A_{\theta}} P$ equipped with operators $\hat W(g) :=W(g)\otimes W^{P}(g)$. 
Analogously one defines a mapping in the opposite direction.

Given a pair $(P, \{W^{P}(g) ; g\in G\})$ as above one can construct a true $(A_{\theta}\rtimes_{R}G , A_{\hat \theta}\rtimes_{R}G)$ 
Morita  equivalence bimodule  in the sense of general definition given at the beginning of this section. 
The construction goes as follows. To shorten notations denote $B_{\theta} = A_{\theta}\rtimes_{R}G$, 
$B_{\hat \theta} = A_{\hat \theta}\rtimes_{R}G$. 
Elements of the $B_{\theta}, B_{\hat \theta})$  bimodule 
that we denote $Q$ are  pairs $p\otimes_{\rm C} \sum_{g} c(g) g$ where $p\in P$ and $\sum_{g} c(g) g$ are formal linear 
combinations of elements of the group $G$ with complex coefficients. Multiplication by complex numbers on $Q$ is defined in 
the obvious way. 
We further define left and right actions of generators of $B_{\theta}$ and  $B_{\hat \theta}$ as 
\begin{eqnarray}
&& U_{\bf n}(p\otimes g) = (U_{R^{-1}(g){\bf n}}p)\otimes g \, , \quad W(h)(p\otimes g) = p\otimes (hg) \, , \nonumber \\
&& (p\otimes g)\hat U_{\bf n} = (p\hat U_{\bf n})\otimes g \, , \quad (p\otimes g)\hat W(h) = (pW^{P}(h))\otimes (gh) \, .
\end{eqnarray}    
One can check that $Q$ satisfies the defining properties of Morita equivalence bimodule, i.e. 
$$
\bar Q\otimes_{B_{\theta}} Q \cong B_{\hat \theta} \, , \qquad    Q\otimes_{B_{\hat \theta}} \bar Q \cong B_{ \theta} \, .
$$
Namely, let fix us the isomorphisms $I: \bar P\otimes_{A_{\theta}} P \to A_{\hat \theta}$, 
$\bar I:   P\otimes_{A_{\hat \theta}} \bar P \to A_{ \theta}$, then one can define 
a mapping   $J: \bar Q\otimes_{B_{\theta}} Q \to B_{\hat \theta}$   as  
$$
J \Bigl( (\bar p \otimes \bar g)\otimes_{B_{\theta}} (p\otimes g) \Bigr) = 
I\Bigl( \bar p \otimes (pW^{P}(g^{-1}\bar g)) \Bigr)\cdot (\bar g^{-1}g) \, .
$$ 
It is easy to check that $J$ is an isomorphism of the corresponding $ (B_{\hat \theta}, B_{\hat \theta})$ bimodules. 
An isomorphism $\bar J:  Q\otimes_{B_{\hat \theta}} \bar Q \to B_{ \theta}$ can be constructed in a similar way.

We further need to construct a correspondence between equivariant connections  on $B_{\theta}$ and $B_{\hat \theta}$ 
modules $F$, $\hat F$. We call an  $A_{\theta}$-connection $\nabla_{i}$ defined on a $B_{\theta}$-module $E$ 
equivariant if it satisfies a constraint 
\begin{equation} \label{equiv}
W^{-1}(g)\nabla_{i}W(g) = R^{j}_{i}(g)\nabla_{j} \, .
\end{equation}
Now let us assume  that we are given a triple $(P,  \nabla^{P},W^{P}(g))$ where $P$ is a 
 $(A_{\theta},A_{\hat \theta})$ Morita equivalence bimodule, operators $W^{P}(g)$ acting on $P$ satisfy (\ref{WP}), 
 $\nabla^{P}_{i}$ satisfies (\ref{biconnect}) and an additional equivariance constraint (\ref{equiv}).
The couple $(P, \nabla^{P})$ establishes a complete Morita equivalence of the algebras $A_{\theta}$. 
It is simple to check that due to condition (\ref{equiv}) a mapping of  $A_{\theta}$- connections that is defined by this 
couple preserves the equivariance condition. We will say that a triple $(P, \nabla^{P}, W^{P}(g))$ specifies a 
$(B_{\theta}, B_{\hat \theta})$ complete equivalence. 
If a triple $(P, \nabla_{\alpha}^{P}, W^{P}(g))$ specifies a $(B_{\theta}, B_{ \theta'})$ complete 
equivalence and a triple  $( P',  \nabla_{\alpha}^{P'},  W^{P'}(g))$ specifies a 
$(B_{ \theta'}, B_{\theta''})$ complete equivalence, then the tensor product 
$ P'' = P\otimes_{B_{\theta'}}  P'$ along with the appropriate connection $ \nabla_{\alpha}^{P''}$ and 
involution $  W^{P''}$ determines $(B_{\theta}, B_{ \theta''})$ complete equivalence. This means that we 
can consider a groupoid of equivalences.

Let us discuss now the existence of triples $(P,  \nabla^{P},W^{P}(g) )$. 
First we would like to note that not every (complete) Morita equivalence of tori can be lifted 
to orbifolds. There is an obvious constraint that the matrix $\theta$ should stay invariant under the orbifold 
group action. Morita equivalence of $A_{\theta}$ algebras is governed by the group $SO(d,d|{\bf Z})$. 
An element $g\in SO(d,d|{\bf Z})$ can be represented by a $2d\times 2d$ block matrix 
$$
g=\left( 
\begin{array}{cc}
M&N\\
R&S
\end{array}
\right)
$$
and the action on $\theta$ is given by a fractional transformation
\begin{equation}\label{fractr}
g: \theta \mapsto (M\theta + N)(R\theta + S)^{-1} \, .
\end{equation}
The orbifold group $G$ naturally acts on $\theta$ as 
$$
h: \theta \mapsto R^{t}(h) \theta R(h) \, , \enspace h\in G \, .
$$
The condition that two actions commute specifies a certain subgroup of $SO(d,d|{\bf Z})$ and 
can be  formalized as follows. We can embed both groups into $O(d,d|{\bf R})$. For the 
first group the embedding is obvious, for the orbifold group we embed an element specified by the matrix $R(g)$ 
into 
$$
R(g) \mapsto \left( 
\begin{array}{cc}
R^{t}(g)&0\\
0&R^{t}(g^{-1})
\end{array}
\right) \, .
$$  
Now the commutation condition has a precise meaning. We proceed to the construction of  $(P,  \nabla^{P},W^{P}(g) )$ triples. 
The group $SO(d,d|{\bf Z})$ is generated by the following transformations. 
The first type of transformations is a subgroup $SL(d,{\bf Z})$ of modular transformations. The second type consists 
of shifts $\theta \mapsto \theta + N$ where $N$ is an arbitrary antisymmetric matrix with integer entries. 
To generate the whole  $SO(d,d|{\bf Z})$  one has to  add a ``flip'' transformation $\sigma$ that inverts 
a $2\times2$ nondegenerate block in the matrix $\theta$. Namely, without loss of generality we can assume that 
$\theta$ has a block form
$$
\theta = \left( 
\begin{array}{cc}
\theta_{11}&\theta_{12}\\
\theta_{21}&\theta_{22}
\end{array}
\right) \, .
$$  
where $\theta_{11}$ is a $2p\times 2p$ nondegenerate matrix. Then a flip $\sigma_{2p}$ sends $\theta$ into 
\begin{equation} \label{flip}
\sigma_{2}(\theta ) =  \left( 
\begin{array}{cc}
\theta_{11}^{-1}&-\theta_{11}^{-1}\theta_{12}\\
\theta_{21}\theta_{11}^{-1}&\theta_{22} - \theta_{21}\theta_{11}^{-1}\theta_{12}
\end{array}
\right) \, .
\end{equation}
One can check that modular transformations, shifts and the flip $\sigma_{2}$ generate the whole $SO(d,d|{\bf Z})$.
For any such transformation that commutes with $G$ one can construct a triple $(P, \nabla^{P}, W^{P}(g))$. 
Let $A \in SL(d,{\bf Z})$ and $A$ commute with $G$. The corresponding $(A_{\theta}, A_{A^{t}\theta A}$ equivalence 
bimodule consists of elements $a\in A_{\theta}$ and the actions of generators  are 
defined as 
\begin{equation} \label{UU}
 U^{l}_{\bf n} (a) = U_{\bf n}a \, , \qquad (a)U_{\bf n}^{r} = aU_{A\bf n} \, .
\end{equation}
where to avoid confusion we denoted by $U_{\bf n}$ elements of $A_{\theta}$, and by $ U^{l}_{\bf n}$, $U_{\bf n}^{r}$ left 
and right actions of the corresponding tori. 
This bimodule can be completed to a triple by adding the following operators $\nabla_{i}^{P}$ and $W^{P}(g)$
\begin{equation} \label{WW}
W^{P}(g) (a) = \rho_g(a) \, , \qquad \nabla^{P}_{i}(a) = \delta_{i}(a)
\end{equation}
where $\rho_g$ stands for automorphisms of algebra $A_{\theta}$ induced by the action 
$R(g)$ on the lattice.

Let $N =(N^{ij})$ be an antisymmetric $G$-invariant matrix with integer entries. The corresponding $(A_{\theta}, A_{\theta + N})$ 
equivalence bimodule consists of elements $a\in A_{\theta}$ with tori actions defined as 
\begin{equation}
U_{\bf n}^{l}(a) = U_{\bf n}a \, , \qquad (a)U^{r}_{\bf n}= aU_{\bf n}(-1)^{\sum_{i<j}n_{i}N^{ij}n_{j}} \, .
\end{equation}  
The operators  $\nabla_{i}^{P}$, $W^{P}(g)$ are the same as in (\ref{UU}), (\ref{WW}).

Finally we need to define a triple corresponding to a G-invariant flip (\ref{flip}). Although it is not hard 
if only somewhat lengthy to describe triples corresponding to generic $G$ invariant flips of the type $\sigma_{2}$ 
we will confine ourselves to the ``total'' flip that inverts the whole matrix $\theta$ (provided the last one is 
invertible). The formulas for that case are most succinct and elegant. Besides this case alone will be 
sufficient for all our future needs.  
The operators $\nabla^{P}_{j}$ should   satisfy 
\begin{equation}\label{Heis}
[\nabla^{P}_{j}, \nabla_{k}^{P}] = -2\pi i \theta_{jk} \cdot {\bf 1}\, .
\end{equation}
Since $\theta$ is nondegenerate the last relation defines a Heisenberg algebra. 
The $(A_{\theta}, A_{\theta^{-1}})$ bimodule is modeled on $L_{2}({\bf R}^{d/2})$ space that is  
assumed to carry an irreducible representation of the Heisenberg algebra (\ref{Heis}). 
The tori actions are defined to be given by the following operators on   $L_{2}({\bf R}^{d/2})$
\begin{equation} \label{L2}
U^{l}_{j} = exp(\nabla^{P}_{j}) \, , \qquad U^{r}_{j} = exp((\theta^{-1})^{jk}\nabla^{P}_{k}) \, .  
\end{equation}
Now the matrix $\theta$ is assumed to be $G$-invariant. This assumption implies that 
 the  transformation $\nabla_{j}\mapsto R_{j}^{k}\nabla^{P}_{k}$ 
is an isomorphism of the Heisenberg algebra (\ref{Heis}). There exists then a set of unitary operators 
$W^{P}(g)$ satisfying (\ref{equiv}). It follows from (\ref{equiv}) and (\ref{L2}) that  the second equation in (\ref{WP})
holds. As for the first equation, in general it is 
satisfied up to a phase factor, i.e. $W^{P}(g)$ determine a projective representation of $G$. 
This possibility can be easily embedded into a general theory discussed above. However we will 
assume that the first equation in (\ref{WP}) is satisfied precisely which is always the case for 
cyclic orbifolds. This finishes the construction of the corresponding triple.

Now we specialize to the cases of ${\bf Z}_{2}$ and ${\bf Z}_{4}$ orbifolds. 
We will show that by  use of Morita equivalence any module admitting a 
constant curvature equivariant connection can be mapped to a module with a nondegenerate 
curvature tensor (the dimension $d$  is assumed to be even).
First consider ${\bf Z}_{2}$ orbifolds. 
In this case the mapping $x \mapsto -x$ always preserves the lattice $D$ and any antisymmetric 
two-form $\theta$ on it. Moreover any transformation from the group $SO(d,d|{\bf Z})$ 
commutes with the ${\bf Z}_{2}$ transformation. In general under transformation (\ref{fractr}) 
the curvature tensor $f = (f_{ij})$ transforms as 
\begin{equation}
\tilde f = AfA^{t} + RA^{t}
\end{equation}
where $A = R\theta + S$. The matrix $\tilde f$ is  nondegenerate if the matrix 
$f + A^{-1}R$ is nondegenerate. One can check that there exist $SO(d,d |{\bf Z})$ transformations 
such that the matrix $A^{-1}R$ is invertible and its matrix norm can be made arbitrarily large. 
It follows from this  that $f + A^{-1}R$ can be made nondegenerate.

In the ${\bf Z}_{4}$ case  we can choose a basis in which a ${\bf Z}_{4}$ generator has a form
$$
\left(
\begin{array}{cc}
0&  1_{2\times 2} \\
-1_{2\times 2} & 0
\end{array} \right)
$$
where for simplicity we set  $d=4$. An arbitrary ${\bf Z}_{4}$ invariant matrix then has a form 
$$
\left( \begin{array}{cc}
A& B \\
-B&A
\end{array} \right)
$$  
where $A$, $B$ are arbitrary $2\times 2$ matrices. In particular we see that we 
have a big supply of invariant matrices with integral entries. This allows one to repeat 
the argument made above for the ${\bf Z}_{2}$ case.

\end{document}